\documentclass[11pt,a4paper]{article}
\usepackage[utf8]{inputenc}
\usepackage[T1]{fontenc}
\usepackage{mathptmx} 
\usepackage[scaled=0.92]{helvet} 
\usepackage[a4paper,margin=25mm]{geometry}
\usepackage{amsmath,amssymb}
\usepackage{graphicx}
\usepackage{booktabs}
\usepackage{array}
\usepackage{xcolor}
\usepackage{setspace}
\usepackage[font=small,labelformat=empty,justification=raggedright,singlelinecheck=false]{caption}
\usepackage[most]{tcolorbox}
\usepackage{lineno}
\usepackage[super,sort&compress,comma]{natbib}
\usepackage[colorlinks=true,linkcolor=black,citecolor=blue!60!black,urlcolor=blue!60!black]{hyperref}
\newcommand{\rf}[1]{ref.~\citenum{#1}}
\newcommand{\rfn}[1]{\citenum{#1}}

\makeatletter
\renewcommand\section{\@startsection{section}{1}{\z@}{2.2ex plus .6ex minus .2ex}{0.6ex}{\normalfont\sffamily\bfseries\large\raggedright}}
\renewcommand\subsection{\@startsection{subsection}{2}{\z@}{1.6ex plus .4ex minus .2ex}{0.3ex}{\normalfont\sffamily\bfseries\normalsize\raggedright}}
\makeatother
\newcommand*\patchAmsMathEnvironmentForLineno[1]{%
  \expandafter\let\csname old#1\expandafter\endcsname\csname #1\endcsname
  \expandafter\let\csname oldend#1\expandafter\endcsname\csname end#1\endcsname
  \renewenvironment{#1}%
    {\linenomath\csname old#1\endcsname}%
    {\csname oldend#1\endcsname\endlinenomath}}
\AtBeginDocument{\patchAmsMathEnvironmentForLineno{equation}\patchAmsMathEnvironmentForLineno{align}}

\newcommand{\dn}{d_{\mathrm{n}}}
\newcommand{\Deff}{D_{\mathrm{eff}}}
\newcommand{\ws}{w_{\mathrm{s}}}

\newcommand{\up}{$^{\uparrow}$}
\newcommand{\dw}{$^{\downarrow}$}

\newtcolorbox{naturebox}[2][]{enhanced,float=tp,colback=black!5,colframe=black!5,boxrule=0pt,arc=0pt,
  left=4mm,right=4mm,top=3mm,bottom=3mm,before skip=5mm,after skip=5mm,
  fonttitle=\sffamily\bfseries\normalsize,coltitle=black,colbacktitle=black!5,
  title={#2},titlerule=0.4pt,titlerule style={black!70},toptitle=3mm,bottomtitle=1.5mm,#1}

\begin{document}
\raggedbottom

\begin{flushleft}
{\sffamily\small\bfseries Perspective}\\[2mm]
{\sffamily\bfseries\LARGE Chasing the KPZ exponents\par}
\vspace{5mm}
{\sffamily\normalsize
Luciano C. Lapas$^{1}$, M\'arcio S. Gomes-Filho$^{2}$ \& Fernando A. Oliveira$^{2,\ast}$\par}

\vspace{3mm}
{\small
$^{1}$Latin-American Institute of Life and Natural Sciences, Federal University for Latin-American Integration, Foz do Igua\c{c}u, Brazil.
$^{2}$International Center of Physics, Institute of Physics, University of Bras\'{\i}lia, Bras\'{\i}lia, Brazil.
$^{\ast}$e-mail: faooliveira@gmail.com\par}
\end{flushleft}
\vspace{2mm}

\setstretch{1.5}

\noindent\textbf{Forty years after it was written down, the Kardar--Parisi--Zhang (KPZ) equation is the paradigm
of scale invariance far from equilibrium. In one dimension it has been solved exactly and confirmed by precision
experiments, and its scaling has now been observed in two-dimensional polariton condensates. Above one dimension,
however, not a single exponent is known exactly, and even the existence of an upper critical dimension is
disputed. In this Perspective we argue that firm statements need not wait for exact solutions. Treating the
nonlinearity as an effective noise that acts on a fractal subset of the substrate leads to bounds on the
roughness exponent, $1/(d+1)\le\alpha\le 1/d$ in $d$ dimensions. We examine the assumptions behind these
bounds, confront them with four decades of theory, simulations and the new experiments, and identify the
tests---foremost in three and more dimensions---that would establish or refute them.}

\vspace{3mm}
\noindent In 1986, Kardar, Parisi and Zhang proposed a continuum description of a growing interface\cite{Kardar86}
that has since become the standard model of scale invariance far from equilibrium (Fig.~1).  For the height
$h(\mathbf{x},t)$ above a $d$-dimensional substrate, the Kardar--Parisi--Zhang (KPZ) equation reads
\begin{equation}
\frac{\partial h}{\partial t}=\nu\nabla^{2}h+\frac{\lambda}{2}\left(\nabla h\right)^{2}+\eta(\mathbf{x},t),
\label{eq:kpz}
\end{equation}
where the first term smooths the interface, the second makes it grow along its local normal and $\eta$ is a
Gaussian white noise of strength $D$. The same equation governs directed polymers in random media and randomly
stirred fluids\cite{HalpinHealyZhang95,Krug97,Takeuchi18}, and it has resurfaced in the phase dynamics of
driven-dissipative condensates\cite{Sieberer25} and in spin transport in quantum magnets\cite{Wei22}. Its
universality class is to nonequilibrium physics what the Ising class is to equilibrium.


\begin{figure}[t]
\centering
\includegraphics[width=\textwidth]{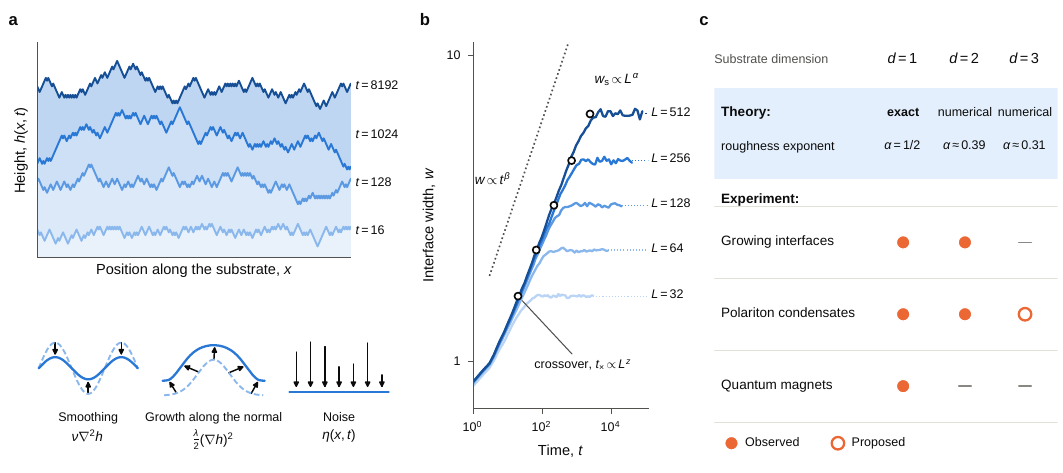}
\caption{\setstretch{1.1}\textbf{Fig.~1 $|$ The Kardar--Parisi--Zhang class at a glance.} \textbf{a}, A growing
interface. Top, height profiles $h(x,t)$ of a simulated interface at four times $t$ (single-step lattice model on a
one-dimensional substrate of 256 sites; profiles are displaced vertically for clarity). Bottom, the three terms of
equation~(\ref{eq:kpz}): surface tension smooths the interface, growth along the local normal adds a term in
$(\nabla h)^{2}$ and the noise $\eta$ roughens it. \textbf{b}, Dynamic scaling. The width $w$ of the interface, here for
the same model with $L=32$ to 512 sites, grows as $t^{\beta}$ and saturates at $\ws\propto L^{\alpha}$ after a crossover
time $t_{\times}\propto L^{z}$ (circles), which defines the roughness, growth and dynamic exponents $\alpha$, $\beta$ and
$z$. The dotted line has the exact one-dimensional slope $\beta=1/3$. \textbf{c}, Theory and experiment, dimension by
dimension. The exponents are known exactly only for $d=1$; for $d=2$ and 3 the values shown are numerical
estimates\cite{PagnaniParisi15,Marinari00}. KPZ scaling has been observed in growing interfaces for
$d=1$ (\rf{TakeuchiSano10}) and $d=2$ (refs~\rfn{Ojeda00},\rfn{Almeida14},\rfn{HalpinHealyPalasantzas14}), in polariton
condensates for $d=1$ (\rf{Fontaine22}) and $d=2$ (\rf{Widmann26}), with a proposal for $d=3$ (\rf{Cao26}), and in
quantum magnets for $d=1$ (\rf{Wei22}).}
\label{fig:schematic}
\end{figure}

In one dimension the theory is a triumph. The scaling exponents have been known exactly from the outset; in 2010, the full distributions of the height fluctuations, which depend on the geometry of the interface, were derived\cite{SasamotoSpohn10,Amir11} and observed in turbulent liquid
crystals\cite{TakeuchiSano10}; and KPZ scaling has since been measured in a polariton condensate\cite{Fontaine22}
and in the spin dynamics of a quantum gas\cite{Wei22}.

In 2026 the experimental frontier moved up a dimension. Measuring the coherence of polariton condensates in
two-dimensional lattices, Widmann and colleagues found a roughness exponent $\alpha=0.42\pm0.03$ for a square
lattice and $0.39\pm0.03$ for a triangular one\cite{Widmann26,Kavokin26}, in agreement with earlier evidence
from the growth of thin films\cite{Ojeda00,Almeida14,HalpinHealyPalasantzas14} and with the value
$\alpha\approx0.39$ long known from simulations\cite{HalpinHealyTakeuchi15}. A three-dimensional polariton
crystal has already been proposed\cite{Cao26}, and neural-network integrators now keep the continuum equation
stable deep in the strong-coupling regime, where conventional schemes diverge\cite{SongXia25}.

Theory has not kept pace. Above one dimension there is no exact solution, either for the exponents or for the
height distributions\cite{HalpinHealy12}, and no controlled expansion; nor is there agreement on whether the
exponents become trivial above some upper critical dimension. Forty years on, the exponents of the simplest
nonequilibrium field theory, in the dimensions that matter for surfaces and for bulk matter, are known only as
decimals extracted from simulations.

In this Perspective we argue that sharp statements need not wait for an exact solution. We advocate a geometric
reading of the KPZ nonlinearity, in which the noise that effectively drives the interface acts on a fractal
subset of the substrate\cite{GomesFilho21RINP,GomesFilho24}. This reading turns the roughness exponent into a
dimension, and the elementary requirement that this dimension lie between $d-1$ and $d$ into bounds on the
exponent. In equilibrium critical phenomena, inequalities between exponents were established well before the
renormalization group explained them; we suggest that KPZ growth is at a comparable stage. Our position is deliberately exposed. One bound is close to a geometric necessity and is respected by all reliable data. The
other is a hypothesis that present simulations probe, and may already challenge, in four dimensions and
above. Both are falsifiable, and we set out how.

\section{One exponent is missing}

\subsection{Scaling relations}

The roughness of an interface of lateral size $L$ is measured by the width $w$ of its height distribution,
which grows as $t^{\beta}$ and saturates at $\ws\sim L^{\alpha}$ after a time of order $L^{z}$
(\rf{Barabasi95}). Matching the two regimes gives $z=\alpha/\beta$, and the invariance of
equation~(\ref{eq:kpz}) under an infinitesimal tilt of the interface---the Galilean invariance of the
equivalent Burgers equation---gives $\alpha+z=2$ in any dimension\cite{Kardar86,Krug97}. A single independent
exponent remains, which we take to be $\alpha$.

In one dimension a fluctuation--dissipation theorem fixes it. The stationary state of
equation~(\ref{eq:kpz}) is that of the linear equation, a random walk in space, so that $\alpha=1/2$,
$\beta=1/3$ and $z=3/2$. For $d\ge2$ the theorem is lost. The renormalization-group flow runs to a
strong-coupling fixed point that perturbation theory cannot reach\cite{Krug97}, and for $d>2$ a roughening
transition separates this rough phase from a smooth one.

\subsection{Three sources of knowledge in higher dimensions}

What is known in higher dimensions comes from three sources. Simulations of lattice models and of directed
polymers\cite{HalpinHealyTakeuchi15} give $\alpha=0.3869(4)$ in two dimensions (\rf{PagnaniParisi15}),
and $0.3135(15)$ and $0.2537(8)$ in three and four (refs~\rfn{Marinari00},\rfn{PagnaniParisi13}). Closed
formulae have been proposed\cite{Wolf87,Kim89,Lassig98,OliveiraTJ22}: $\alpha=1/(d+1)$ by Wolf and Kert\'esz,
$2/(d+3)$ by Kim and Kosterlitz, the quantized values $2/5$ and $2/7$ for $d=2$ and 3 by L\"assig and, most
recently, $7/(4d+10)$ by T.~J.~Oliveira. The non-perturbative renormalization group, the analytical method
that most directly reaches the strong-coupling fixed point, yields $\alpha\approx0.37$ in two dimensions but $0.18$
in three\cite{Kloss12,Canet25}. Finally, mode-coupling and field-theoretic arguments place an upper critical
dimension, above which $\alpha=0$, at $d_{\mathrm{c}}\le4$ (refs~\rfn{LassigKinzelbach97},\rfn{Colaiori01}),
whereas real-space arguments\cite{Castellano98} and simulations, now carried up to $d=15$
(refs~\rfn{OliveiraTJ22},\rfn{Alves14}), find none.

\subsection{What rigorous mathematics guarantees}

Rigorous mathematics, decisive in one dimension, has so far constrained higher dimensions only weakly. For
first-passage percolation, a lattice member of the class, the available inequalities and scaling
relation\cite{Kesten93,Licea96,Chatterjee13} amount to no more than $0\le\alpha\le2/3$ in any dimension, and
the two-dimensional equation has been controlled only in weak-noise regimes, far from the strong-coupling
fixed point\cite{Caravenna20}. A window that wide does not even exclude an upper critical dimension.

\section{A dimension for the noise}

\subsection{An exact fluctuation--dissipation balance}

Our starting point is an exact rewriting. Subtracting from the nonlinear term its mean, which merely shifts
the growth velocity, leaves a linear equation driven by an effective noise (Box~1). The difficulty has not
disappeared: it now resides in this noise, which is correlated in space and time and is shaped by the
interface that it drives. In the stationary state its amplitude at each wavevector balances the smoothing
exactly, so that a fluctuation--dissipation relation survives in every dimension, albeit with a noise
strength $\Deff$ that depends on scale\cite{GomesFilho21EPL,GomesFilho25}. In one dimension $\Deff$ is
constant and the equilibrium result is recovered. In higher dimensions it grows towards large scales, and the
rate of that growth is the roughness exponent in disguise.

\begin{naturebox}{Box 1 $|$ From effective noise to roughness exponent}
\setstretch{1.15}\small
\textbf{An exact rewriting.} Let $v=(\lambda/2)\langle(\nabla h)^{2}\rangle$ be the mean contribution of the
nonlinear term to the growth velocity. Equation~(\ref{eq:kpz}) is identical to
\begin{equation}
\frac{\partial h}{\partial t}=\nu\nabla^{2}h+v+\phi,\qquad \phi=\eta+\psi,\qquad
\psi=\frac{\lambda}{2}\left(\nabla h\right)^{2}-v .
\label{eq:rewrite}
\end{equation}
This is an Edwards--Wilkinson equation with a constant drift, driven by an effective noise $\phi$ of zero
mean. Unlike $\eta$, $\phi$ is correlated in space and in time---its correlation time at scale $\ell$ grows as
$\ell^{z}$---and it depends on the interface.

\smallskip
\textbf{A fluctuation--dissipation balance at every scale.} For a stationary interface on a lattice of $N$
sites, with structure factor $S(k)=\langle|\hat h_{\mathbf{k}}|^{2}\rangle/N$, It\^o's lemma applied to
$|\hat h_{\mathbf{k}}|^{2}$ gives, for $\mathbf{k}\neq0$,
\begin{equation}
\nu k^{2}S(k)=\Deff(k)\equiv D+\frac{1}{N}\,\mathrm{Re}\,\langle\hat\psi_{\mathbf{k}}\hat h_{-\mathbf{k}}\rangle .
\label{eq:balance}
\end{equation}
The relation is exact for any $\lambda$ and $d$. In one dimension the stationary measure is Gaussian, the cubic
correlation vanishes and $S(k)=D/\nu k^{2}$, hence $\alpha=1/2$. For $d\ge2$, $S(k)\sim k^{-(d+2\alpha)}$
implies $\Deff(k)\sim k^{-\theta}$ with $\theta=d+2\alpha-2>0$ ($\theta\approx0.77$ for $d=2$): the theorem
holds scale by scale, with an amplitude that grows towards large scales.

\smallskip
\textbf{The noise dimension.} The KPZ coupling $\lambda^{2}D/\nu^{3}$ has the dimension of
(length)$^{d-2}$. For $d\ge2$, power counting therefore cannot supply a positive roughness exponent, which must
come from an anomalous dimension. The proposal\cite{GomesFilho24} assigns it to the noise: the effective noise acts on a
set of dimension $\dn\le d$, so that, with heights measured as lengths, $\Deff/\nu$ has the dimension of
(length)$^{\dn}$ instead of (length)$^{d}$. If, as in one dimension, where $\ws^{2}=DL/12\nu$, the saturated
width depends on the system size only through the product $\Deff L/\nu$, then
\begin{equation}
\ws\propto\left(\frac{\Deff L}{\nu}\right)^{1/(\dn+1)},
\label{eq:ws}
\end{equation}
which is equation~(\ref{eq:alpha}) of the main text. Galilean invariance then gives
$z=(2\dn+1)/(\dn+1)$ and $\beta=1/(2\dn+1)$.

\smallskip
\textbf{Status.} Equations~(\ref{eq:rewrite}) and (\ref{eq:balance}) are exact. Equation~(\ref{eq:ws}) rests on
the single-combination hypothesis, which is exact only for $d=1$; at present $\dn$ is inferred from $\alpha$ and
not measured independently.
\end{naturebox}

\subsection{The noise as a fractal object}

How should such a noise be characterized? Fractal geometry is intrinsic to the problem: a self-affine
interface with roughness exponent $\alpha$ has fractal dimension $d_{\mathrm{f}}=d+1-\alpha$ (\rf{Barabasi95}), and nothing
obliges the noise that sculpts it to be any less fractal. Lattice models of the KPZ class make the point
concrete. In the single-step and restricted solid-on-solid models, sites are visited at random---this is the
white noise---but the growth rule is applied or not according to the local shape of the interface. A rejected
attempt is noise that is never realized. What survives acts on a subset of the substrate selected by the interface
itself, and for a self-affine interface there is no reason for that subset to have an integer dimension. The
hydrodynamic picture points the same way. In the Burgers equation, for which $-\nabla h$ is a velocity,
dissipation concentrates on shocks, which in $d$ dimensions are fronts of dimension $d-1$
(\rf{BecKhanin07}); a random forcing multiplies and roughens them.

We therefore attribute to the effective noise a fractal dimension $\dn$ (\rf{GomesFilho24}). If, as in one
dimension, the system size enters the saturated width only through the combination $\Deff L/\nu$, dimensional
analysis gives (Box~1)
\begin{equation}
\alpha=\frac{1}{\dn+1}.
\label{eq:alpha}
\end{equation}
For $d=1$, $\dn=1$ returns $\alpha=1/2$: the noise fills the substrate, which is the geometric content of the
fluctuation--dissipation theorem. For $d\ge2$ the most precise numerical estimates translate into
$\dn\approx1.58$, $2.19$ and $2.94$ in two, three and four dimensions. The support of the noise becomes ever
sparser relative to the substrate.

\subsection{Status of the relation}

Taken by itself, equation~(\ref{eq:alpha}) is a change of variable from $\alpha$ to $\dn$. Its content lies in
what $\dn$ is asserted to be: the dimension of a set. In this respect it resembles the Fisher exponent $\eta$ of
equilibrium critical phenomena, which can be read as the codimension of the fractal set that supports the
critical correlations\cite{Lima24}.

\section{Two bounds}

\subsection{Derivation}

A set on which the noise acts must satisfy two requirements. It is contained in the substrate, so that
$\dn\le d$. And the space in which the effective dynamics unfolds---the support of the noise together with the
height direction, of dimension $\dn+1$---must be able to contain the $d$-dimensional substrate, so that
$\dn\ge d-1$. The first limit, $\dn=d$, is a noise that, like $\eta$, acts everywhere on the substrate; the
second, $\dn=d-1$, is a noise concentrated on a set of codimension one, lines for $d=2$ and surfaces for
$d=3$---in the hydrodynamic picture, dissipation carried by shock fronts rather than by their intersections.
Equation~(\ref{eq:alpha}) then confines the roughness exponent to
\begin{equation}
\frac{1}{d+1}\le\alpha\le\frac{1}{d},
\label{eq:bounds}
\end{equation}
and the dynamic exponent to $2-1/d\le z\le2-1/(d+1)$ (Fig.~2). The lower bound on $\alpha$ is the
Wolf--Kert\'esz formula, which appears here not as a conjecture for the exponent but as a limit that it cannot
cross. It is attained in one dimension.

\subsection{Unequal standing}

The two bounds do not have the same standing. The lower one requires only that the support of the effective
noise be a subset of the substrate. The upper one adds a hypothesis about how sparse that support can become,
and it is there that the proposal is most exposed.

\subsection{A narrow band}

The band defined by equation~(\ref{eq:bounds}) is narrow. Its width, $1/[d(d+1)]$, decreases faster than the
exponent itself, so that its midpoint fixes $\alpha$ to within a fraction $1/(2d+1)$: $\pm20\%$ in two
dimensions and $\pm14\%$ in three. That is enough to discriminate between proposals (Table~1). The
Kim--Kosterlitz formula leaves the band for $d>3$ and that of T.~J.~Oliveira for $d>10/3$. The
renormalization-group value in three dimensions falls below the band, as would any theory with a finite upper
critical dimension. Inside the band lie the simulations in two and three dimensions and every experiment
performed so far (Fig.~3).

\begin{figure}[t]
\centering
\includegraphics[width=\textwidth]{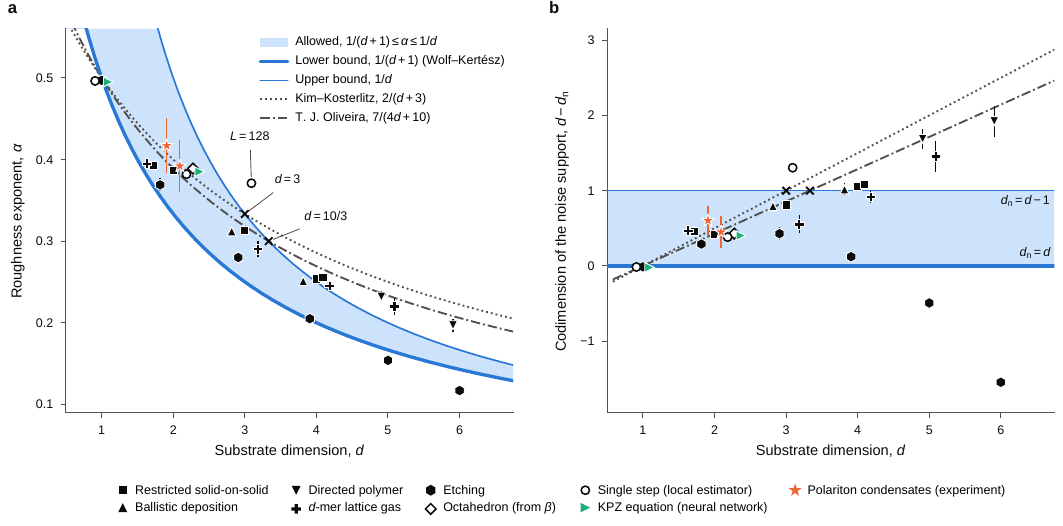}
\caption{\setstretch{1.1}\textbf{Fig.~2 $|$ Fractal bounds on the roughness exponent.} \textbf{a}, Roughness
exponent $\alpha$ as a function of the substrate dimension $d$. The shaded band is the region allowed by
equation~(\ref{eq:bounds}), between the lower bound $1/(d+1)$ (thick line),
which coincides with the Wolf--Kert\'esz formula\cite{Wolf87}, and the upper bound $1/d$ (thin line), continued to non-integer
$d$. The formulae of Kim and Kosterlitz\cite{Kim89} and of T.~J.~Oliveira\cite{OliveiraTJ22} cross the upper
bound at $d=3$ and $d=10/3$ (crosses). Symbols are published estimates, displaced horizontally within each
dimension for clarity; the most precise stationary estimate is plotted at the integer value of $d$. Filled black symbols, roughness measured in the stationary state of restricted
solid-on-solid\cite{Marinari00,PagnaniParisi13,PagnaniParisi15}, ballistic-deposition\cite{AlvesFerreira16},
directed-polymer\cite{Kim19}, $d$-mer\cite{Odor10} and etching\cite{Rodrigues15} models; open symbols, the
octahedron model\cite{Kelling18} (converted from the growth exponent) and the single-step
model\cite{GomesFilho24} (local estimator; the point labelled $L=128$ is not converged); triangles,
neural-network integration of the KPZ equation\cite{SongXia25}; stars, polariton condensates\cite{Widmann26} on
a square (upper) and a triangular (lower) lattice. Error bars are the published uncertainties and are often
smaller than the symbols. \textbf{b}, The same data expressed as the codimension of the support of the effective
noise, $d-\dn=d+1-1/\alpha$. The allowed region is the strip between $\dn=d$ and $\dn=d-1$.}
\label{fig:bounds}
\end{figure}

\begin{table}[t]
\caption{\setstretch{1.1}\textbf{Table 1 $|$ The roughness exponent between two and six dimensions: bounds,
proposals and simulations}}
\label{tab:scorecard}
\small\setstretch{1.1}
\begin{tabular*}{\textwidth}{@{\extracolsep{\fill}}lccccc@{}}
\toprule
 & $d=2$ & $d=3$ & $d=4$ & $d=5$ & $d=6$\\
\midrule
\multicolumn{6}{@{}l}{\textbf{Bounds, equation~(\ref{eq:bounds})}}\\
Lower bound, $1/(d+1)$ & 0.333 & 0.250 & 0.200 & 0.167 & 0.143\\
Upper bound, $1/d$ & 0.500 & 0.333 & 0.250 & 0.200 & 0.167\\
\addlinespace
\multicolumn{6}{@{}l}{\textbf{Proposals}}\\
Kim--Kosterlitz\cite{Kim89}, $2/(d+3)$ & 0.400 & 0.333 & 0.286\up & 0.250\up & 0.222\up\\
L\"assig\cite{Lassig98} & 0.400 & 0.286 & -- & -- & --\\
T.~J.~Oliveira\cite{OliveiraTJ22}, $7/(4d+10)$ & 0.389 & 0.318 & 0.269\up & 0.233\up & 0.206\up\\
Non-perturbative renormalization group\cite{Kloss12} & 0.373(1) & 0.179(4)\dw & 0.075(4)\dw$^{,\ast}$ & -- & --\\
\addlinespace
\multicolumn{6}{@{}l}{\textbf{Simulations}}\\
Restricted solid-on-solid\cite{PagnaniParisi15,Marinari00,PagnaniParisi13} & 0.3869(4) & 0.3135(15) & 0.2537(8)\up & -- & --\\
Ballistic deposition\cite{AlvesFerreira16} & -- & 0.312(2) & 0.251(5) & -- & --\\
$d$-mer lattice gas\cite{Odor10} & 0.395(5) & 0.29(1) & 0.245(5) & 0.22(1)\up & --\\
Directed polymer\cite{Kim19} & -- & -- & -- & 0.232(7)\up & 0.197(8)\up\\
Etching\cite{Rodrigues15} & 0.369(8) & 0.280(7) & 0.205(3) & 0.154(2)\dw & 0.117(1)\dw\\
From growth exponents\cite{OliveiraTJ22} & 0.390(3) & 0.314(6) & 0.264(9) & 0.225(8)\up & 0.197(8)\up\\
\bottomrule
\end{tabular*}
\par\vspace{1.5mm}
{\footnotesize Numbers in parentheses are the published uncertainties in the last digits. \up\ (\dw), above the
upper (below the lower) bound, by at least two standard deviations where an uncertainty is quoted.
$^{\ast}$Leading-order approximation. The last row converts the growth exponents $\beta$ averaged over several
models in \rf{OliveiraTJ22} with $\alpha=2\beta/(1+\beta)$.\par}
\end{table}

\section{No upper critical dimension}

\subsection{Consequence of the lower bound}

The most consequential implication is also the simplest. Because $\alpha\ge1/(d+1)>0$, the interface is rough,
and its dynamics faster than diffusive ($z<2$), in every finite dimension. There is no upper critical
dimension. The exponent vanishes only as $d\to\infty$, in keeping with the flat interfaces obtained when the
equation is integrated on fully connected graphs\cite{MarcosMerino26}. This conclusion rests on the more
robust of the two bounds. It sides with the
simulations\cite{Marinari00,PagnaniParisi13,Alves14,AlvesFerreira16,Kim19,OliveiraTJ22} and against the
predictions of $d_{\mathrm{c}}\le4$ (refs~\rfn{LassigKinzelbach97},\rfn{Colaiori01}). It also has a cost.
Without an upper critical dimension there is no mean-field theory to expand around, which is one reason why
inequalities, rather than expansions, may be the natural tool for this problem.

\subsection{Why four dimensions may still be special}

Four dimensions may nevertheless be special, for a different reason. The most precise estimate
there\cite{PagnaniParisi13}, $\alpha=0.2537(8)$, corresponds to $\dn=2.94(1)$, within 2\% of $d-1$ (Fig.~2b). In the
present language, $d=4$ is not where roughness ends but where the support of the effective noise has thinned
to codimension one. It is tempting to speculate that this structural change is what approximate theories
register as a critical dimension.

\section{What the simulations say}

\subsection{The lower bound}

How do the bounds fare against four decades of numerical work? The lower one fares well. Below it we find only
the etching-model estimates\cite{Rodrigues15} for $d=5$ and 6, which differ by 30--40\% from those of other
models in the same dimensions\cite{Odor10,Kim19} and therefore cannot be asymptotic. Every other estimate,
whether measured in the stationary state or converted from growth exponents compiled up to $d=15$
(\rf{OliveiraTJ22}), respects it by a wide margin.

\subsection{The upper bound}

The upper bound is another matter. It holds comfortably in two and three dimensions. In four it is saturated:
three estimates\cite{Marinari00,Odor10,AlvesFerreira16} agree with $1/4$ within two standard deviations, and
the most precise\cite{PagnaniParisi13} exceeds it by 1.5\%, or 4.6 standard deviations. In five and six
dimensions, directed-polymer simulations\cite{Kim19} give $\alpha=0.232(7)$ and $0.197(8)$, which are 16\% and
18\% above $1/d$, and the growth exponents of restricted solid-on-solid models\cite{Alves14,OliveiraTJ22}
continue this trend up to $d=15$, where they imply $\alpha=0.099(7)$ against $1/d=0.067$.

\subsection{Two readings and how to decide}

Two readings are possible. In the first, the estimates for $d\ge5$ have not reached the asymptotic regime. The
lattices are small---linear sizes of at most 32 in the stationary state for $d=5$ (\rf{Kim19}), and of 4 to 6
for $d\ge12$ (\rf{OliveiraTJ22})---and finite-size effects can displace an estimate by more than the width of
the band: our own single-step simulation in three dimensions\cite{GomesFilho24}, analysed with a local estimator
on a lattice of size $L=128$, overshoots the upper bound by 11\%, whereas the estimates from larger lattices lie
inside (Fig.~2). 
Even in four dimensions, the formula $7/(4d+10)$, which describes the growth exponents well, lies 6\% above the
stationary estimate\cite{PagnaniParisi13}. Against this reading stands the mutual agreement of independent
models and of a real-space renormalization scheme\cite{OliveiraTJ22}.

In the second reading the estimates are right, and above four dimensions the support of the effective noise
acquires a codimension larger than one---dissipation migrating, in the hydrodynamic picture, from shock
fronts to their intersections. The data are then described by a noise dimension that grows linearly,
$\dn\approx1+0.58(d-1)$---in essence the formula $7/(4d+10)$, for which $\dn=(4d+3)/7$---and only the lower
bound survives. The two readings make different predictions, and
the difference can be resolved: matching in five dimensions the $4\times10^{9}$ sites already simulated in
four\cite{PagnaniParisi13} would raise the linear size from 32 to more than 80. Because corrections to scaling decay roughly as the
inverse of the linear size in three and four dimensions\cite{Marinari00,PagnaniParisi13}, that should be enough
to see whether the excess over $1/d$ shrinks.

\section{Two dimensions as a test bed}

\subsection{Experiments}

Two dimensions is where experiment, simulation and theory now meet (Fig.~3). The band there is
$1/3\le\alpha\le1/2$, or $1/5\le\beta\le1/3$. For a condensate, whose coherence decays as a stretched
exponential with exponent $2\alpha$ in space and $2\beta$ in time\cite{Widmann26}, this means exponents
between $2/3$ and 1, and between $2/5$ and $2/3$. Every measurement lies inside the band: the two polariton
lattices\cite{Widmann26}, silica and oligomer films\cite{Ojeda00,HalpinHealyPalasantzas14} and, for $\beta$,
CdTe films\cite{Almeida14}. The experimental uncertainties, however, are 7--9\% for $\alpha$ and larger for
$\beta$, whereas the candidate values lie within a few per cent of one another. To discriminate between them,
experiments must become five to ten times more precise. The long coherence times and tunable geometry of
polariton lattices make that a realistic goal.

\begin{figure}[t]
\centering
\includegraphics[width=\textwidth]{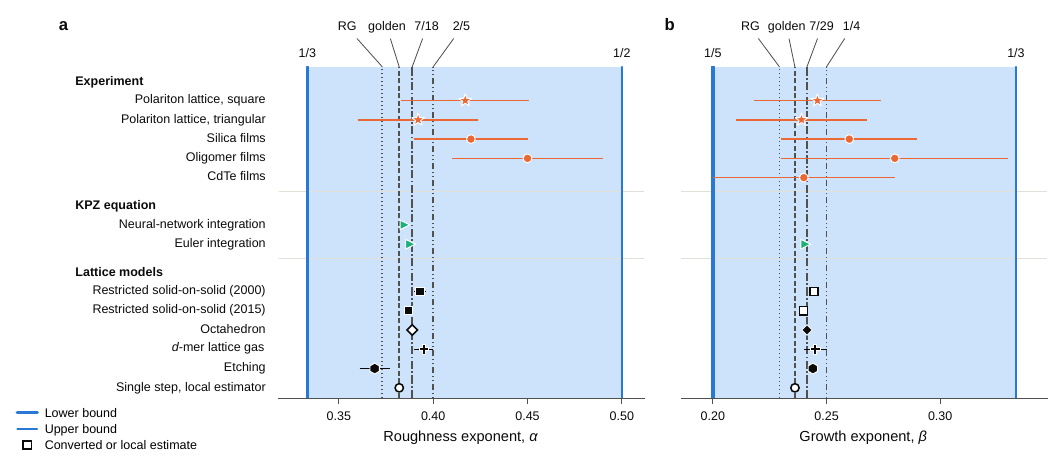}
\caption{\setstretch{1.1}\textbf{Fig.~3 $|$ Two dimensions as a test bed.} \textbf{a},\textbf{b}, Roughness
exponent $\alpha$ (\textbf{a}) and growth exponent $\beta$ (\textbf{b}) for $d=2$. Experiments: polariton
lattices\cite{Widmann26}, silica films grown by chemical vapour deposition\cite{Ojeda00}, oligomer
films\cite{HalpinHealyPalasantzas14} and CdTe films\cite{Almeida14}. Integrations of the KPZ equation: neural
network\cite{SongXia25} and Euler scheme\cite{HalpinHealy12}. Lattice models: restricted
solid-on-solid\cite{Marinari00,PagnaniParisi15}, octahedron\cite{Kelling18}, $d$-mer\cite{Odor10},
etching\cite{Rodrigues15} and single-step\cite{GomesFilho24}. Filled symbols are values reported in the original
work; open symbols are obtained from the other exponent through $z=\alpha/\beta$ and $\alpha+z=2$, or from a
local estimator. Error bars are the published uncertainties. The shaded region is the band of
equation~(\ref{eq:bounds}). Vertical lines mark candidate values: RG, non-perturbative renormalization
group\cite{Kloss12}; golden, $\alpha=(3-\sqrt5)/2$ (refs~\rfn{GomesFilho21RINP},\rfn{GomesFilho24}); $7/18$ and
$7/29$, T.~J.~Oliveira\cite{OliveiraTJ22}; $2/5$ and $1/4$, Kim--Kosterlitz\cite{Kim89} and
L\"assig\cite{Lassig98}.}
\label{fig:2d}
\end{figure}

\subsection{Simulations}

Simulations already discriminate. They exclude the value $2/5$ of Kim--Kosterlitz and L\"assig by more than
thirty standard deviations\cite{PagnaniParisi15}, and they are divided over $7/18$, which agrees with the value
inferred from the growth exponent\cite{Kelling18} but lies five standard deviations above the most precise
stationary estimate\cite{PagnaniParisi15}.

\subsection{A curious regularity}

Simulations also bear on a curious regularity. In two dimensions a level set of the interface has dimension
$2-\alpha$ and the support of the noise has dimension $\dn=1/\alpha-1$, and the two almost
coincide: for $\alpha=0.3869$ they are 1.61 and 1.58. Were they equal, $\alpha$ would be
$(3-\sqrt5)/2\approx0.382$ and $z$ the golden mean\cite{GomesFilho21RINP,GomesFilho24}. A local estimator
applied to single-step simulations returns this value to four digits\cite{GomesFilho24}; but the same estimator
is still 0.6\% below the exact value at the end of the corresponding one-dimensional run, and the global
estimates\cite{PagnaniParisi15,Kelling18} are 1.3--1.8\% higher. We therefore regard the golden mean as the
zeroth order of an expansion in the mismatch between the two dimensions, not as an exact result. The mismatch
is small only for $d=2$---it is 0.5 in three dimensions---and why it should be so is an open and, we think,
significant question.

\section{Outlook}

Exact solutions of the KPZ equation above one dimension may be a long way off. Bounds are a more modest goal,
and we have argued that a geometric view of the noise supplies them: a lower bound on the roughness exponent,
which excludes an upper critical dimension and is respected by all reliable data, and an upper bound, which
holds in the dimensions accessible to experiment, is saturated in four and is contested beyond.

Three developments would turn this position into a result, or dispose of it. The first is numerical.
Simulations in four to six dimensions with linear sizes several times those available today, based on
stationary observables and with an explicit treatment of corrections to scaling, would decide whether
$\alpha\le1/d$ survives. Stable integrators of the continuum equation\cite{SongXia25} offer a check that does
not depend on any lattice model. The second is conceptual. At present the noise dimension is inferred from
$\alpha$. An observable that measured the support of the effective noise directly would promote $\dn$ from a
reparametrization to a prediction; the scale-dependent amplitude $\Deff(k)$, accessible in simulations and,
through the structure factor, in experiments, is the natural place to look. A relation between $\dn$ and the
fractal dimension of the interface would then close the problem, as the analogous relation does for the
Fisher exponent\cite{Lima24}. The third is experimental: a measurement of $\alpha$ in two dimensions at the
1\% level, and any measurement at all in three, where the bounds $1/4\le\alpha\le1/3$, $1/7\le\beta\le1/5$ and
$5/3\le z\le7/4$ await the polariton crystals that have been proposed\cite{Cao26}.

The KPZ equation has repeatedly rewarded the belief that simple structure lies behind its difficulty. That its
exponents should be confined, in every dimension, between two elementary fractions would be such a structure.
Forty years after the equation was written down, the claim can be put to the test.

\nolinenumbers
\setstretch{1.15}
\bibliographystyle{naturemag}
\bibliography{references}

\small
\section{Data availability}
All exponents shown in Figs~2 and 3 and in Table~1 are taken from the cited publications; Fig.~1a,b shows simulations of the single-step model performed for this article. The compiled values
are available from the corresponding author upon request.

\section{Code availability}
The scripts that compile the data and generate the figures are available from the corresponding author upon
request.

\section{Use of AI tools}
A large language model (Claude, Anthropic) was used to assist in editing the text and in compiling the
published values from the cited literature; no image was produced by a generative image model. The authors have checked the text, the compiled values, the scripts and the references against the original sources and take full responsibility for the content.

\section{Acknowledgements}
This work was supported by CNPq (grants APV-0064-1.05/25 and 303119/2022-5 to F.A.O.). M.S.G.-F.
acknowledges the National Laboratory for Scientific Computing (LNCC/MCTI, Brazil) for computing resources on
the SDumont supercomputer. L.C.L. acknowledges support from the National Institute of Science and Technology in
Innovative Research in Health Sciences---from Nanotechnology to Artificial Intelligence (INCT PICS), CNPq,
grant no.~408417/2024-2.

\section{Author contributions}
All authors contributed to the ideas discussed here and to the writing of the manuscript.

\section{Competing interests}
The authors declare no competing interests.

\section{Additional information}
\textbf{Correspondence} should be addressed to Fernando A. Oliveira.

\end{document}